\documentclass[prb,preprint,amsmath,amssymb]{revtex4-1}

\usepackage{graphicx}
\usepackage{dcolumn}
\usepackage{bm}

\begin{document}

\title{Intrinsic Terahertz Plasmons and Magnetoplasmons in Large Scale Monolayer Graphene}

\author{I. Crassee$^{1}$, M. Orlita$^{2,3}$, M. Potemski$^{2}$, A. L. Walter$^{4,5}$, M. Ostler$^{6}$,
Th. Seyller$^{6}$, I. Gaponenko$^{1}$, J. Chen$^{7,8}$, and A. B.
Kuzmenko$^{1}$}

\affiliation{
\\
\mbox{ }
\\
$^{1}$D\'epartement de Physique de la Mati\`ere Condens\'ee, Universit\'e de
Gen\`eve, CH-1211 Gen\`eve 4,
Switzerland\\
\\
$^{2}$Grenoble High Magnetic Field Laboratory, CNRS-UJF-UPS-INSA F-38042 Grenoble Cedex 09, France\\
\\
$^{3}$Faculty of Mathematics and Physics, Charles University, Ke Karlovu 5, 121 16 Praha 2, Czech Republic\\
\\
$^{4}$Departement of Molecular Physics, Fritz-Haber-Institut der
Max-Planck-Gesellschaft, Faradayweg 4-6, 14195 Berlin,
Germany\\
\\
$^{5}$E. O. Lawrence Berkeley National Laboratory, Advanced
Light Source, MS6-2100, Berkeley, CA 94720\\
\\
$^{6}$Lehrstuhl f\"{u}r Technische Physik, Universit\"{a}t
Erlangen-N\"{u}rnberg, Erwin-Rommel-Str. 1, 91058 Erlangen, Germany
\\
$^{7}$CIC nanoGUNE Consolider, 20018 Donostia-San Sebastian, Spain\\
\\
$^{8}$Centro de Fisica de Materiales (CSIC-UPV/EHU) and Donostia
International Physics Center (DIPC), 20018 Donostia-San Sebastian, Spain}


\date{\today}

\begin{abstract}
\textbf{We show that in graphene epitaxially grown on SiC the Drude
absorption is transformed into a strong terahertz plasmonic peak due to
natural nanoscale inhomogeneities, such as substrate terraces and wrinkles.
The excitation of the plasmon modifies dramatically the magneto-optical
response and in particular the Faraday rotation. This makes graphene a unique
playground for plasmon-controlled magneto-optical phenomena thanks to a
cyclotron mass 2 orders of magnitude smaller than in conventional plasmonic
materials such as noble metals.}

\end{abstract}

\maketitle

Graphene attracts a lot of attention as a novel optoelectronic and plasmonic
material for applications ranging from the terahertz to the
visible\cite{WunschNJP06, HwangPRB07, MikhailovPRL07, RanaIEEE08,
JablanPRB08, XiaNatureNanoTech09, BonaccorsoNaturePhoton10, BludovEPL10,
EchtermeyerNatureComm11, JuNatureNano11, KoppensNanoLett11, FeiNanoLett11,
EnghetaScience11, NikitinPRB11}. One expects that plasmon waves in graphene
can be squeezed into a much smaller volume\cite{KoppensNanoLett11,
EnghetaScience11} than in noble metals routinely used in plasmonics and,
importantly, can be manipulated by external gate voltage. A Dirac-like linear
electronic dispersion and zero bandgap in graphene make its electromagnetic
response rather unusual compared to other known two-dimensional conductors,
such as 2D electron gases (2DEGs) in semiconductor
heterostructures\cite{AndoRevModPhys82}, even though the basic description of
the propagating plasma modes is essentially the same\cite{WunschNJP06,
HwangPRB07}.

In order to observe plasmonic absorption optically one generally has to break
the translational invariance of the system. Typical ways to couple plasmons
to electromagnetic radiation in two-dimensional systems are placing an
external grid in the vicinity of the sample\cite{AllenPRL77} and making
stripe- or dotlike periodic structures inside the system\cite{AllenPRB83,
HeitmannSurfSci92, KukushkinPRL03}. In graphene, this coupling has recently
been achieved by pattering it in the shape of ribbons\cite{JuNatureNano11}
and by using a metallic atomic force microscopy (AFM) tip in scattering-type
scanning near field optical microscopy\cite{FeiNanoLett11}. However, no
measurement of graphene plasmons in magnetic field was reported so far.

Graphene epitaxially grown on SiC shows a number of intrinsic uniformly
distributed defects caused by the substrate terraces and thermal relaxation
after the graphitization process\cite{CambazCarbon08, EmtsevNatureMat09,
CamaraPRB09, FortiPRB11}. In this Letter, we demonstrate that these defects,
usually considered a nuisance, are in fact beneficial to excite terahertz
plasmons in graphene without the need for artificial structuring. Because of
the small cyclotron mass of the Dirac-like charge carriers in graphene we
observe a strong effect of the magnetic field on the plasma modes,
reminiscent of the behavior observed in 2DEGs. This is in drastic contrast to
conventional plasmonic materials like gold, where the carrier mass exceeds
the free-electron mass, leading to at best a weak dependence on magnetic
field\cite{SepulvedaPRL10}.

A layer of honeycomb carbon was grown on the silicon side of SiC by a
graphitization procedure at 1450 $^\circ$C in an Argon atmosphere, as
described in our previous work\cite{CrasseeNatPhys11}. Subsequent hydrogen
passivation of the silicon dangling bonds transformed the so-called buffer
layer into quasi freestanding graphene\cite{RiedlPRL09, SpeckMatSciForum10}.
X-ray photoemission spectroscopy confirmed that the thickness was one
graphene layer and that the backside of the SiC substrate was graphene free.
After hydrogenation, graphene becomes strongly p-doped; the Fermi level,
$\epsilon_F =$ -0.34 eV, is below the Dirac point, which corresponds to a
hole concentration $n=\epsilon_F^2/(\pi\hbar^2 v_F^2)\approx 8 \times
10^{12}$ cm$^{-2}$\cite{CrasseeNatPhys11}, where $v_F \approx 10^6$ m/s is
the Fermi velocity. All magneto-optical measurements were done in the Faraday
geometry (magnetic field and propagation of light normal to the sample) using
a Fourier transform infrared spectrometer connected to a split-coil
superconducting magnet\cite{CrasseeNatPhys11}. The illuminated area of
several square millimeters was fully covered by graphene.

\begin{figure}
    \includegraphics[width=8.5cm]{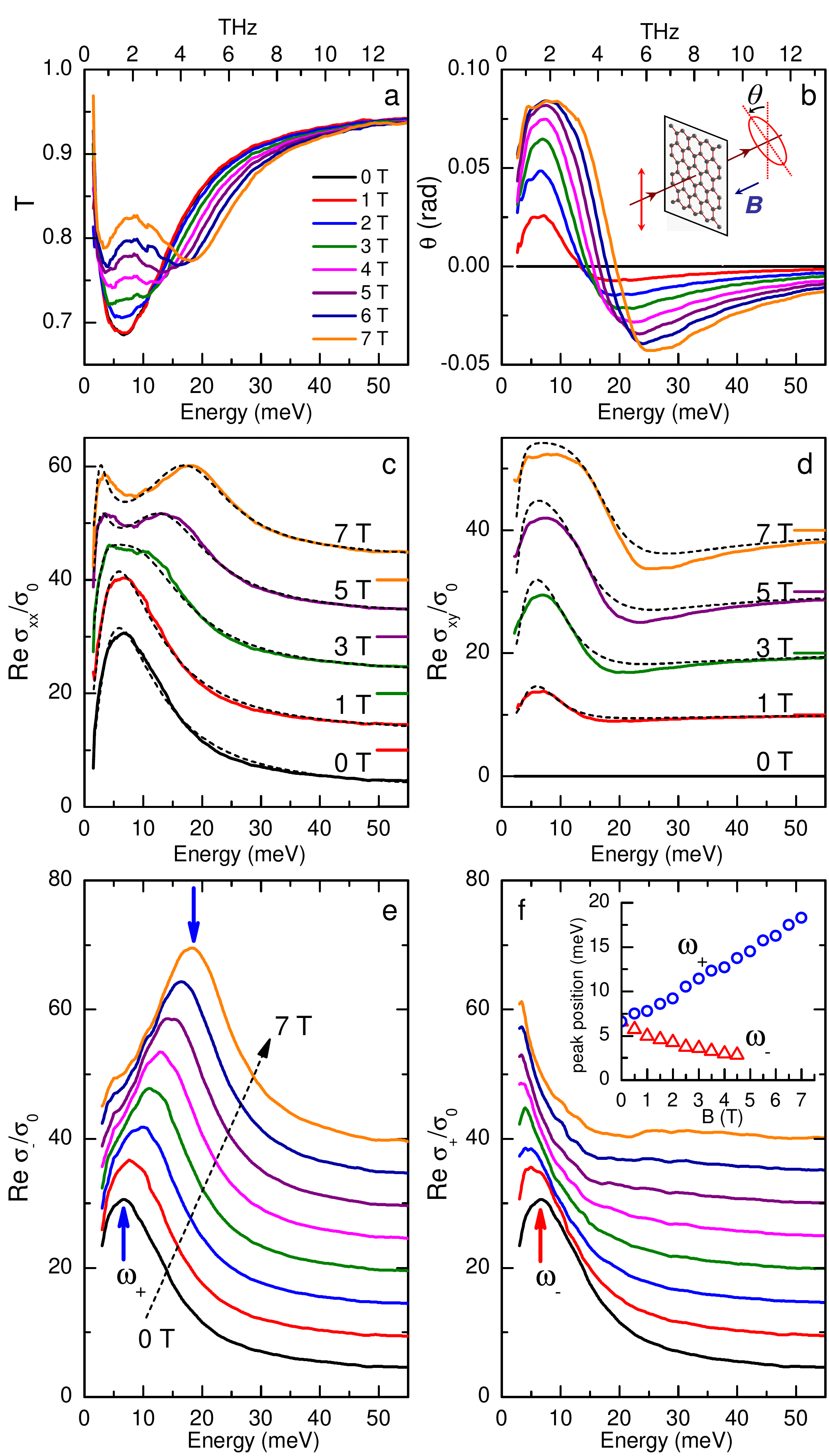}\\
    \caption{(a) Terahertz transmission spectra of graphene on SiC normalized to the
    bare substrate and (b) the Faraday rotation spectra.
    The inset of (b) shows the polarization state of light before and after passing the sample in the Faraday rotation
    experiment.
    The diagonal (c) and Hall (d) conductivities normalized to $\sigma_0 = e^2/4\hbar$.
    The optical conductivity for left (e) and right (f) circularly polarized light.
    The magnetic field dependence of the peak positions $\omega_+$ and $\omega_-$ are shown in the inset of (f).
    The spectra in panels c, d, e, and f are offset for clarity.
    The zero lines in (c) and (d) are indicated by lines of the same color.}
    \label{FigS}
\end{figure}

Figures \ref{FigS}a,b show the optical transmission and Faraday rotation
spectra measured at 5 K in fields $B$ up to 7 T. The diagonal conductivity
$\sigma_{xx}(\omega)$, normalized to the universal conductivity $\sigma_0 =
e^2/4\hbar$\cite{AndoJPSJ02}, shown in Figure \ref{FigS}c is obtained from
the optical transmission, as described in the Supporting Information. At zero
magnetic field, we observe a strong maximum centered around 6.5 meV (1.6 THz)
instead of the normally expected Drude peak at zero frequency due to free
carriers. As we demonstrate below, the deviation from the Drude behavior is
associated with the presence of a confinement potential acting on free
carriers and the corresponding plasmonic absorption. A similar resonance was
observed recently in graphene micro-ribbons in the polarization perpendicular
to the ribbons\cite{JuNatureNano11}.

When a magnetic field is applied, the plasma resonance splits into two modes,
one of which increases and the other decreases with $B$ (Figure \ref{FigS}c).
In order to get further insight into the origin of these modes we extracted
the Hall conductivity $\sigma_{xy}(\omega)$ from the Faraday rotation spectra
(Figure \ref{FigS}d) and obtained the optical conductivity for left and right
circularly polarized radiation, $\sigma_\pm(\omega) = \sigma_{xx}\pm
i\sigma_{xy}$ (Figures \ref{FigS}e,f), as described in the Supporting
Information. One can clearly see that each of the modes is excited only in
one circular polarization. The peak positions in $\sigma_{-}(\omega)$ and
$\sigma_{+}(\omega)$ we denote $\omega_+$ and $\omega_-$ respectively, since
the former increases and the latter decreases with magnetic field. The inset
of Figure \ref{FigS}f shows the field dependence of $\omega_+$ and
$\omega_-$. The field-induced splitting of the plasmon peak resembles
strikingly the appearance of collective resonances observed previously in
disk-shaped quantum dots of two-dimensional electron gases based on GaAs
heterostructures \cite{AllenPRB83,KukushkinPRL03} and in bound 2D electrons
on the surface of liquid helium \cite{MastPRL85,GlattiPRL85}. In both cases,
the upper and lower branches were attributed to the so-called bulk and edge
magnetoplasmons, respectively, with the frequencies\cite{AllenPRB83}
\begin{equation}
    \omega_\pm=\sqrt{\frac{\omega_c^2}{4} + \omega_0^2} \pm \frac{|\omega_c|}{2},
    \label{Eqomegaplusmin}
\end{equation}
\noindent where $\omega_0$ is the plasmon frequency at zero field, $\omega_c
= \pm eB/mc$ is the cyclotron frequency, defined as positive for electrons
and negative for holes, $m$ is the cyclotron mass, and $c$ the speed of
light. At high fields ($|\omega_c|\gg\omega_0$), the upper branch becomes
essentially the usual cyclotron resonance with a linear dependence on
magnetic field, while the lower branch represents a collective mode confined
to the edges\cite{FetterPRB85} with the energy inversely proportional to the
field.
%
\begin{figure*}
    \includegraphics[width=17cm]{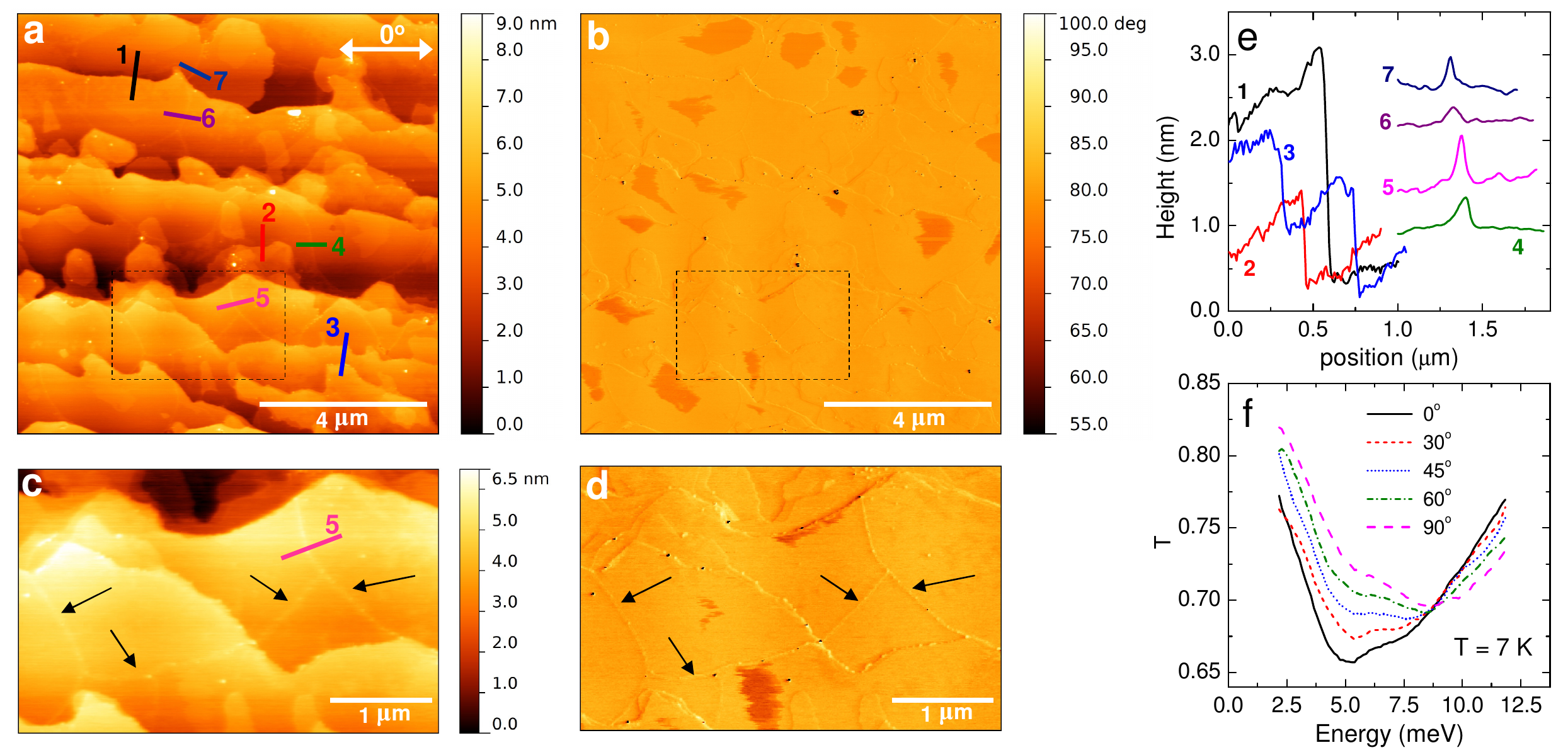}\\
    \caption{Topographic height (a) and phase (b) of a 10$\times$10 $\mu$m$^{2}$ region of epitaxial graphene on SiC used
    for optical experiments.
    (c) and (d) Close-ups of the regions in (a) and (b) marked by
    dashed rectangles. Arrows point to graphene wrinkles. Height profiles (e) along selected lines marked in
    (a), corresponding to terraces (1-3) and wrinkles (4-7).
    Terahertz transmission (f) for different polarizations with respect to the zero direction indicated by the white
    arrow in (a).}
    \label{Figafm}
\end{figure*}

In order to clarify the origin of the confinement that causes the plasmonic
resonance, we performed vibrating cantilever AFM imaging, allowing us to
extract topographic and phase information. In Figure \ref{Figafm}a a
topographical height image of a 10 $\times$ 10 $\mu$m$^2$ area of the sample
is presented. The dominating structures are the terraces due to the miscut
angle of SiC. Their irregular shape as compared to morphologies observed
earlier \cite{EmtsevNatureMat09} is related to the specific graphitization
temperature used in our work. Importantly, the terraces are oriented in the
same direction across the entire sample. Figure \ref{Figafm}b presents the
map of the oscillation phase of the cantilever on the same area. The dark
spots in the phase correspond to regions without graphene, as was determined
by Raman spectroscopy. Closer inspection of the AFM images also reveals
numerous wrinkles such as the ones indicated by the arrows in Figures
\ref{Figafm}c,d. The wrinkles are formed due to the relaxation of strain in
graphene during the cooling down after the
graphitization\cite{CambazCarbon08}. Figure \ref{Figafm}e shows height
profiles for the lines marked in Figure \ref{Figafm}a. Profiles 1 to 3
correspond to steps in the SiC substrate, while traces 4 to 7 are taken
across the wrinkles on the terraces and show that these wrinkles have a
height of less than 1 nm, which is in agreement with previous
work\cite{CambazCarbon08}. The regions of homogeneous graphene have a typical
size of about one micrometer throughout the whole sample.

Polarized optical transmission (Figure \ref{Figafm}f) provides a hint that
the terahertz resonance peak is related to the morphological structures seen
by AFM. A significant anisotropy is found, which correlates with the
orientation of the terraces. In particular, the absorption maximum is at
about 1.7 times higher energy for the electric field perpendicular to the
terraces than for the parallel orientation. Note that the peak position
remains at finite energy for every polarization. The excitation of the
plasmon parallel to the terraces is likely due to the rough shape of the SiC
step edges. One cannot exclude that the wrinkles, which are randomly
oriented, might also play a role in confining the carriers. Because of the
anisotropy, the curves and the plasmon energy in Figure \ref{FigS} are
effective averages over all polarizations.

It is worthwhile to notice that the terahertz transmission of multilayer
graphene epitaxially grown on the carbon side of SiC also reveals an
absorption peak, although at a somewhat lower energy, with a strong
polarization dependence as shown in the Supporting Information. Although it
might have a similar plasmon origin as in the case of monolayer graphene on
the silicon face of SiC, the presence of many layers with different doping
levels makes this interpretation less straightforward.

We found that it is possible to describe quantitatively the plasmon structure
and its splitting in magnetic field by a Drude-Lorentz formula for the
optical conductivity\cite{AllenPRB83,MikhailovPRB96}:
\begin{equation}
\sigma_\pm(\omega) = \frac{2 D}{\pi}\frac{i}{\omega \mp \omega_c + i \gamma -\omega_0^2/\omega },
\label{sigmaplusmin}
\end{equation}
\noindent where $D$ is the plasmon spectral weight and $\gamma$ is the
scattering rate. In the present case, this equation should be regarded as
purely phenomenological, although it can be rigorously derived for certain
types of inhomogeneous media, such as disk-shaped quantum dots, using the
effective medium Maxwell-Garnett approach (the relevant details are given in
the Supporting Information). Equation \ref{sigmaplusmin} is the simplest
analytical expression which reduces to a Lorentzian shape in the limit of
zero field ($\omega_{c} = 0$) and describes the usual Drude cyclotron
resonance when the plasmon energy is vanishing ($\omega_{0} = 0$). For small
values of $\gamma$, it indeed has resonances $\omega_{\pm}$ at frequencies
given by eq. \ref{Eqomegaplusmin}. We fitted the experimental spectra of both
$\sigma_{xx} = (\sigma_+ + \sigma_-)/2$ and $\sigma_{xy} = (\sigma_+ -
\sigma_-)/2\,i$ at every magnetic field to eq. \ref{sigmaplusmin}, treating
$\omega_{0}$, $\omega_{c}$, $\gamma$ and $D$ as adjustable parameters. We
also added a small frequency independent background term $\sigma_b \approx$
3.5 $\sigma_0$ to the real part of the diagonal optical conductivity, which
may have various origins, as discussed in the Supporting Information. The
fits are shown as dashed lines in Figures \ref{FigS}c,d. The experimental
data including all important spectral features are well reproduced, which,
given the complexity of the sample, comes as an encouraging surprise to be
explained in future studies. However, the noticeable deviations, especially
in $\sigma_{xy}(\omega)$, demonstrate the limitations of this simple model
with respect to our sample.

\begin{figure}
    \includegraphics[width=6cm]{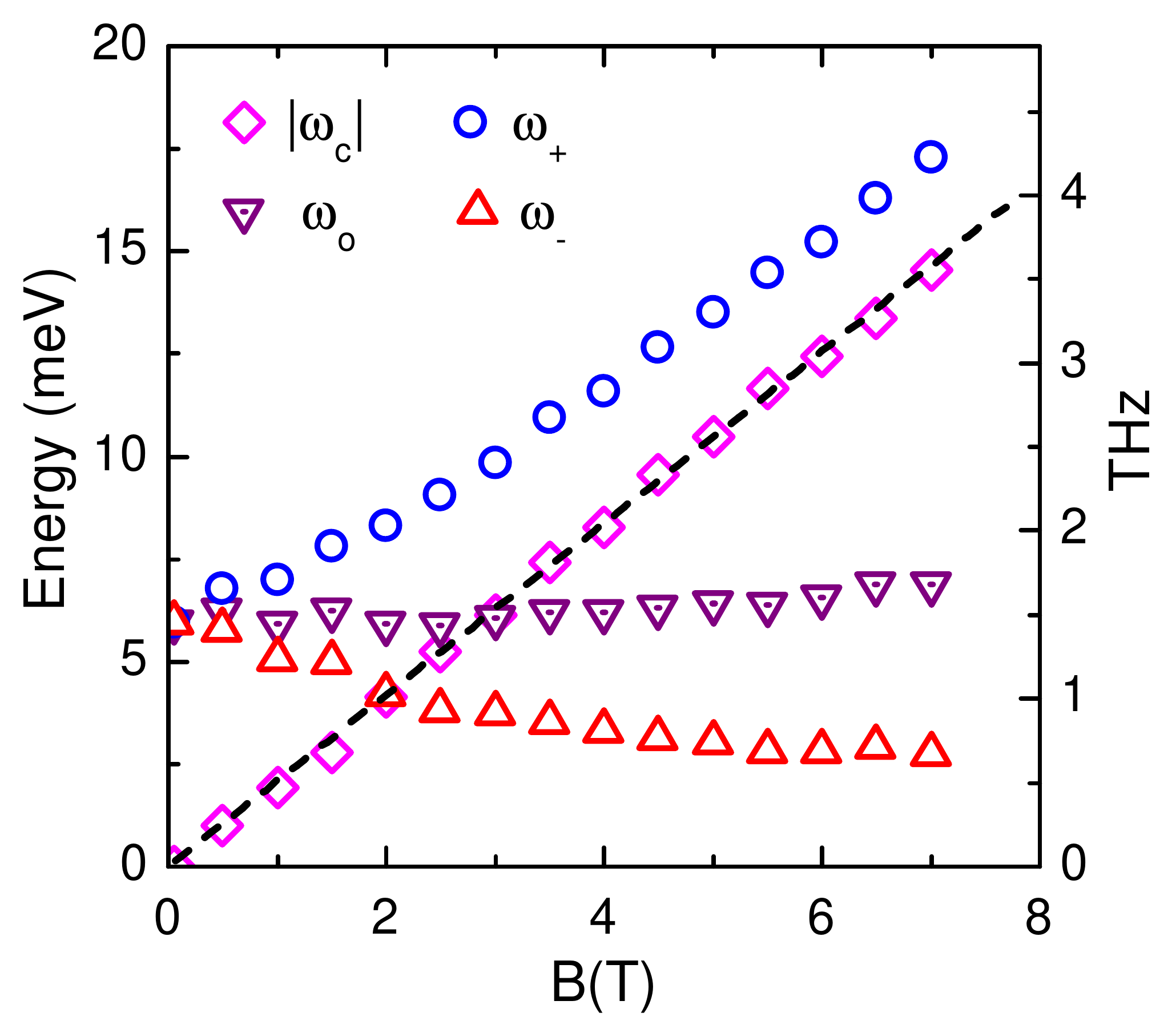}\\
    \caption{The magnetic field
    dependence of the plasmon energy $\hbar\omega_0$, cyclotron
    resonance energy $\hbar\omega_c$ and magnetoplasmon energies
    $\hbar \omega_\pm$. The dotted line is a linear
    fit to the cyclotron energy.}
    \label{FigColorPlot}
\end{figure}

Figure \ref{FigColorPlot} shows the field dependence of $\omega_{0}$ and
$\omega_{c}$ extracted from the fitting procedure. The bare plasmon frequency
is essentially constant, while the cyclotron frequency demonstrates a nearly
perfectly linear growth with a slope $\hbar|\omega_{c}|/B$ = 2.1 meV/T
corresponding to a cyclotron mass of 5.5 \% of the free electron mass $m_e$.
Note that in our previous work\cite{CrasseeNatPhys11} an apparent deviation
from the linear dependence was reported because the plasmonic contribution
was not taken into account. In Figure \ref{FigColorPlot}, we also plot the
magnetoplasmon frequencies $\omega_\pm$ calculated using eq.
\ref{Eqomegaplusmin} and the experimental values of $\omega_c$ and
$\omega_0$. They are very close to the peak positions in $\sigma_{\pm}$
(Figure \ref{FigS}e,f). The spectral weight of the plasmon peak shows only a
small, if any, magnetic field dependence. The value of $\hbar D/\sigma_0$
extracted from the fitting is about 0.52 eV. Taken alone, it is somewhat
smaller than the expected value of $2|\epsilon_F$| = 0.68 eV. The difference
might be related to a noncomplete coverage of the substrate by graphene (see
Figure \ref{Figafm}b), however can also be due to the presence of the
background $\sigma_b$. The broadening, $\hbar\gamma$, of the peak is about
10-12 meV, which is more than two times smaller than observed in a similar
sample by another group\cite{YanACSNano}. It might still be larger than the
intrinsic electron scattering, since the spectral feature is additionally
broadened by the distribution of sizes and shapes of the homogeneous regions.
A more detailed discussion of the spectral weight, background, and scattering
is given in the Supporting Information.

The Dirac-like charge carriers in graphene at high doping are expected to
show a classical cyclotron resonance with a linear dependence on magnetic
field\cite{AndoJPSJ02}, which perfectly agrees with our data. The cyclotron
mass depends on doping according to the relation $m=|\epsilon_F|/v_F^2$.
Using $|\epsilon_F| =$ 0.34 eV, and $m =$ 0.055 $m_e$, we find that $v_F =$
1.04 $\times$ 10$^{6}$ m/s, which matches remarkably well the Fermi velocity
obtained by other methods\cite{OrlitaSemiSciTech10}.

The plasmon frequency, $\hbar\omega_0 = $ 6.5 meV, contains important
information about the intrinsic properties of the electron gas and the
confinement causing the plasmon excitation. It is instructive to compare the
value of $\omega_0$ found in this work with the resonant frequency for the
reference case of isolated disk-shaped quantum dots, for which the effective
medium model predicts\cite{AllenPRB83, LeavittPRB86, MikhailovPRB96}:
\begin{equation}
\omega_{0}^2 = \frac{3\pi^2 n e^2}{2 m d \kappa},
\label{Eqomega0}
\end{equation}
\noindent where $\kappa = (1 + \epsilon_{SiC})/2 \approx 5$ is the average
dielectric constant of the surrounding media and $d$ is the dot diameter. For
$d = 1$ $\mu$m, which roughly corresponds to the mean size of homogeneous
regions in our sample, and for the same charge density $n = 8 \times 10^{12}$
cm$^{-2}$, the expected plasmon frequency would be 15.2 meV. This is more
than twice the experimentally observed value. Lacking a rigorous quantitative
model for the present sample, we ascribe this difference to the fact that the
defect lines that separate homogeneous graphene regions are very narrow, and
that the electromagnetic coupling between neighboring regions therefore plays
an important role. Such a coupling causes a redshift\cite{MikhailovPRB96,
RomeroOpticsexpress06, LassiterNanoLett08, NikitinArxiv11} of the plasmon
energy as compared to the case of noncoupled particles, which is commonly
observed in plasmonic nanostructures.

\begin{figure}
    \includegraphics[width=8.5cm]{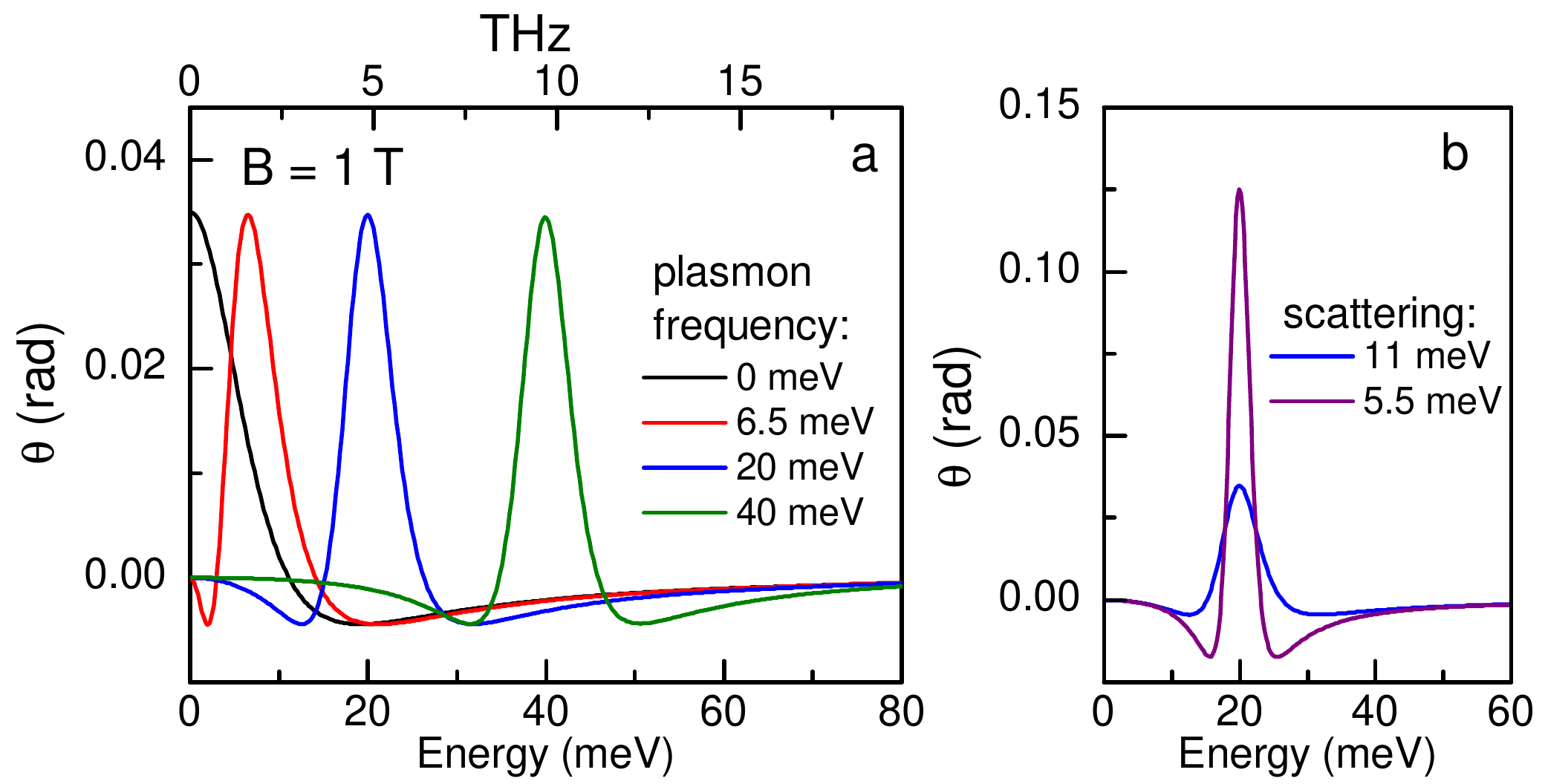}\\
    \caption{A simulation of the effect of the plasmon energy and the scattering on the Faraday rotation.
    (a) Faraday rotation for $\hbar\gamma$ = 11 meV for different values of $\hbar\omega_0$ = 0, 6.5, 20, and 40 meV.
    (b) Faraday rotation for $\hbar\omega_0$ = 20 meV and different values of $\hbar\gamma$ = 11 and 5.5 meV.
    In both cases the magnetic field is 1 T.}
    \label{FigFaradayAngle}
\end{figure}

Important for terahertz applications is the question to what extent the
presence of a plasmon affects the Faraday rotation in graphene. The
simulations in Figure \ref{FigFaradayAngle}a demonstrate that the Faraday
angle is at maximum close to the magnetoplasmon resonance and therefore can
be controlled not only by magnetic field but also by $\omega_0$. Here we
compare the calculated Faraday rotation of homogeneous graphene ($\omega_0 =$
0) and the rotation in the presence of plasmons such as the one observed in
our sample (6.5 meV) and for higher plasmon energies (20 and 40 meV). A way
to increase the plasmon frequency is to decrease the size of the homogeneous
regions, which could be done, for example, by varying the miscut angle of the
substrate. Notice that a route to increase the Faraday rotation for a given
plasmon frequency is to use samples with reduced electronic scattering as
demonstrated in Figure \ref{FigFaradayAngle}b. Rotations above 0.1 radians by
just one atomic layer at a modest field of 1 T do not seem to be out of
experimental reach.

Similarly, in more conventional plasmonic materials the Faraday and Kerr
angles are enhanced close to the plasma resonance\cite{SepulvedaPRL10,
BonanniNanoLett11}. For example, in a recent work a Kerr rotation on the
order of 10$^{-4}$ to 10$^{-3}$ radians was detected in an array of Au
disks\cite{SepulvedaPRL10}. Thus, graphene shows more than 2 orders of
magnitude larger rotation, which is a direct consequence of a much smaller
cyclotron mass even though the carrier density per unit cell is also much
lower than in noble metals. The measurements presented in this Letter were
performed at low temperature to maximally resolve the magnetoplasmonic
spectral structures. We do believe, though, that the magnetoplasmonic
phenomena described here will persist up to room temperature, based on our
temperature dependent experimental study of the cyclotron resonance in
epitaxial graphene on the carbon face of SiC\cite{CrasseePRB11}.

In conclusion, we found that morphological defects on the nanoscale such as
atomic steps in SiC and wrinkles in epitaxial graphene produce a remarkably
strong plasmon resonance. This resonance has essentially the same origin as
the plasmon peak observed in two-dimensional electron gases and in
nanostructured graphene. The important difference, however, is that the
confinement potential in epitaxial graphene is natural and does not require
special lithographic patterning, which risks reducing the carrier mobility.
Instead, one can think of controlling the plasmon frequency by varying the
preparation of the substrate and the graphitization process. The presence of
the plasmon dramatically changes the cyclotron resonance and Faraday
rotation. Graphene appears to be a unique material, where one finds
simultaneously a small effective mass giving rise to strong magneto-optical
effects and excellent plasmonic properties. This combination opens pathways
toward plasmon-controlled terahertz magneto-optics.

After the submission of this paper we became aware of the work by H. Yan \emph{et al.}\cite{YanNatNano12,YanArxiv12}, which reports the observation of plasmons and magnetoplasmons in micron sized graphene disks.

This work was supported by the Swiss National Science Foundation (SNSF) by
Grants 200021-120347 and IZ73Z0-128026 (SCOPES program), through the National
Centre of Competence in Research ``Materials with Novel Electronic
Properties-MaNEP" and by projects EuromagnetII, GACR P204/10/1020 and
GRA/10/E006 (Eurographene-EPIGRAT) and the ESF Eurographene project
``Graphic-RF". We thank R. Hillenbrand, S. A. Mikhailov, and D. van der Marel
for useful discussions.

\newpage
\begin{center}{\bf \large Supporting Information}
\end{center}

\setcounter{equation}{0}
\setcounter{figure}{0}

\subsection{Extraction of the optical conductivity components from transmission and Faraday rotation
spectra}\label{thinfilm}

\begin{figure}
    \includegraphics[width=8cm]{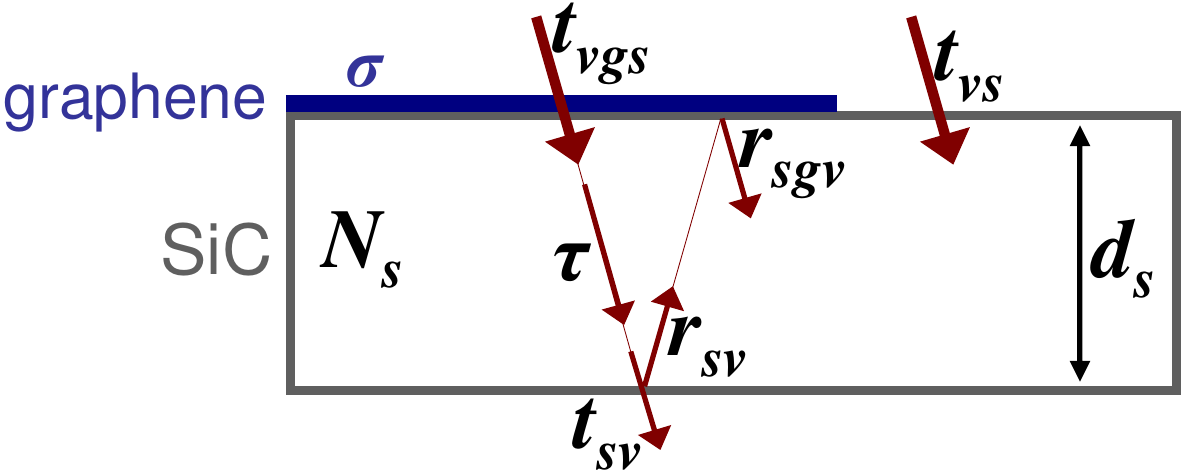}\\
    \caption{\textbf{Supporting Information.} The schematic representation of the sample used
    and the definitions of the complex coefficients of eq. \ref{coefs}.
    A near normal incidence was used, the rays on the figure are inclined for clarity.}
    \label{Fig1}
\end{figure}

In deriving the relations between the optical conductivity of graphene and
the measured spectra, we must take into account multiple reflections in the
substrate, which is partially transparent in the considered spectral range.
One can do the standard plane-wave counting in the ``vacuum ($v$) / graphene
($g$) / substrate ($s$) / vacuum ($v$)" system using the complex coefficients
(Figure \ref{Fig1}):
\begin{eqnarray}
t_{vs} &=& \frac{2}{N_s + 1}, \: t_{sv} = \frac{2N_{s}}{N_s + 1}, \, t_{vgs} = \frac{2}{N_s + 1 + Z_{0}\sigma},\nonumber\\
r_{sv} &=& \frac{N_{s} - 1}{N_s + 1}, \: r_{sgv} = \frac{N_{s} - 1 - Z_{0}\sigma}{N_s + 1+ Z_{0}\sigma}, \: \tau = \exp(i\omega
N_{s}d_{s}/c).
\label{coefs}
\end{eqnarray}
\noindent Here $N_{s} = n_{s} + i k_{s}$ and $d_{s}$ are the experimentally
known complex refractive index and the thickness of the substrate, $Z_{0}$ =
377 $\Omega$ is the impedance of vacuum and $\sigma$ is the two-dimensional
optical conductivity of graphene. In our measurements the spectral resolution
was reduced to 5 cm$^{-1}$ in order to suppress the Fabry-Perot interference.
In this case the multiple reflected rays add incoherently (\emph{i.e.} by
power and not by electric field), and the experimental transmission
coefficients of the bare substrate and the sample are given by:
\begin{eqnarray}
T_{vsv} = \frac{\left|t_{vs}t_{sv}\tau\right|^2}{1 - \left|r_{sv}^2\tau^2\right|^2}, \;
T_{vgsv} =  \frac{\left|t_{vgs}t_{sv}\tau\right|^2}{1 - \left|r_{sgv}r_{sv}\tau^2\right|^2},
\end{eqnarray}
\noindent which results in the substrate-normalized transmission:
\begin{eqnarray}
T &=& \frac{T_{vgsv}}{T_{vsv}} = \left|\frac{t_{vgs}}{t_{vs}}\right|^2\frac{1 - \left|r_{sv}^2\tau^2\right|^2}{1 -
\left|r_{sgv}r_{sv}\tau^2\right|^2}\nonumber \\
&=& \frac{|N_{s} + 1|^4 - |N_{s} - 1|^4|\tau|^4}{|(N_{s} + 1)(N_{s} + 1 + Z_{0}\sigma)|^2 - |(N_{s} - 1)(N_{s} - 1 -
Z_{0}\sigma)|^2|\tau|^4}.
\end{eqnarray}
\noindent Since $k_{s} \ll n_{s}$ in the transparency window of the
substrate, one can safely neglect $k_{s}$ in all complex coefficients (eq.
\ref{coefs}), except in $\tau$. Defining the absorption coefficient of the
substrate $\alpha = |\tau|^2 = \exp(-2\omega k_{s}d_{s}/c)$ and the
substrate-vacuum amplitude reflectivity $r=(n_{s}-1)/(n_{s}+1)$ we arrive,
after some algebra, at the relation:
\begin{equation}
T = \left[1 + \frac{2}{n_{s} + 1}\frac{1 + \alpha^2 r^3}{1 - \alpha^2 r^4}\:\mbox{Re}\left(Z_{0}\sigma\right)
+ \frac{1}{(n_{s} + 1)^2}\frac{1 - \alpha^2 r^2}{1 - \alpha^2 r^4}\left|Z_{0}\sigma\right|^2\right]^{-1}.
\label{Tinv}
\end{equation}
In principle, the power transmission alone is not sufficient to determine the
complex conductivity $\sigma$. However, the prefactor in the second term is
much larger than the one in the third term, therefore the imaginary part of
$\sigma$ affects the transmission much less than the real part. Under these
circumstances a reasonable approximation is to invert the above equation with
respect to $\mbox{Re}(\sigma)$ by setting $\mbox{Im}(\sigma)$ to zero. In
magnetic field, eq. \ref{Tinv} remains strictly valid for each circular
polarization separately, if one substitutes $\sigma$ with $\sigma_{\pm}$. In
unpolarized light, however, the transmission is an average of the two. In
order to extract the diagonal conductivity shown in Figure 1c of the main
text, we made a further simplification by ignoring the relatively small
effect of $\sigma_{xy}$ on the transmission and substituting $\sigma$ with
$\sigma_{xx}$ in eq. \ref{Tinv}.

Likewise, the Faraday rotation is effectively averaged over different
internal reflections in the substrate. Experimentally, the Faraday angle is
determined from the position of the minimum in the optical signal as a
function of the relative angle between two polarizers set before and after
the sample. Adding the substrate reflections incoherently, we obtain the
exact expression:
\begin{equation}
\theta = \frac{1}{2}\mbox{Arg}\frac{t_{vgs,-}t_{vgs,+}^{*}}{1 -
r_{vgs,-}r_{vgs,+}^{*}\left|r_{sv}\tau^2\right|^2},
\end{equation}
\noindent where $t_{vgs,\pm}$ and $r_{vgs,\pm}$ are obtained from eq.
\ref{coefs} by substituting $\sigma\rightarrow\sigma_{\pm}$. Although the
Faraday angle is a function of all components of the optical conductivity
tensor, we found that for the relevant parameter values the simple linear
approximation holds:
\begin{equation}
\theta \approx  \frac{1}{n_{s} + 1}\frac{1 + \alpha^2 r^3}{1 -
\alpha^2 r^4}\:\mbox{Re}\left(Z_{0}\sigma_{xy}\right).
\end{equation}
\noindent This allows us to determine the real part of the Hall conductivity
(Figure 1d of the main text) directly from the Faraday angle. In order to
extract the real part of $\sigma_{\pm} = \sigma_{xx}\pm i\sigma_{xy}$, shown
in Figures 1e and 1f of the main text, the imaginary part of $\sigma_{xy}$
has to be determined as well. For this purpose we used the Kramers-Kronig
relations for $\sigma_{xy}$, using the model curves to extrapolate data
beyond the experimental range.

\subsection{Polarization dependent transmission of multilayer graphene on the carbon side of SiC}\label{EMA}

\begin{figure}
    \includegraphics[width=8cm]{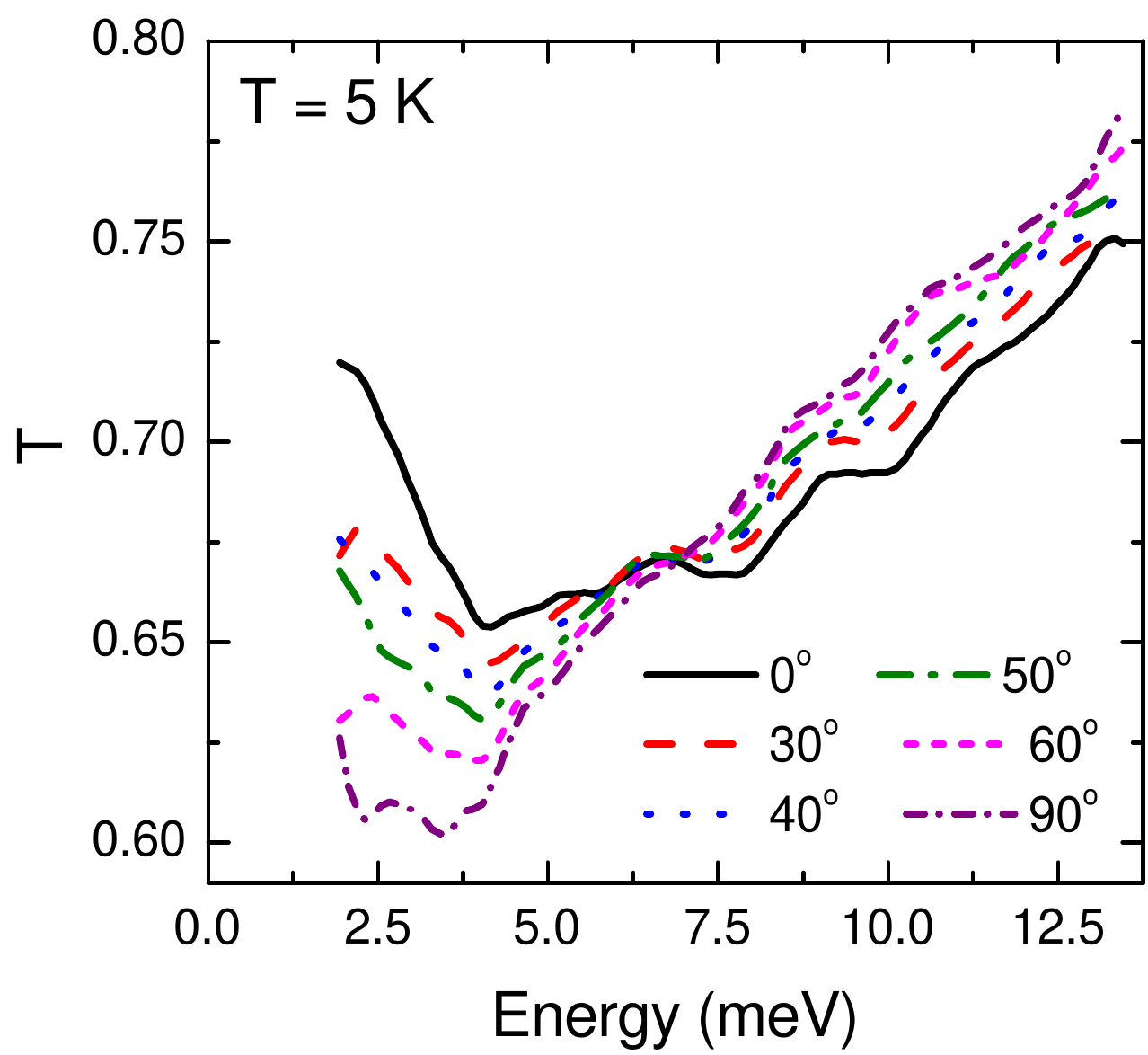}\\
    \caption{\textbf{Supporting Information.} Terahertz transmission of multilayer graphene on the C face of
    SiC for different polarizations at 5 K and 0 T. The zero position of the polarizer is chosen randomly.}
    \label{Fig3}
\end{figure}

The THz transmission spectra of multilayer graphene epitaxially grown on the
carbon side of SiC are shown in Figure \ref{Fig3} for different polarizations
at 5 K and 0 T. The sample has 4 to 6 rotationally stacked layers. Two
observations can be made: (i) the spectra are strongly polarization dependent
and (ii) there is a transmission dip (absorption peak) between 3 and 5 meV
depending on the polarization. Therefore this sample also shows a clear
deviation from the Drude behavior at low enough frequencies, similar to
graphene on the silicon face of SiC discussed in the main text. The peak
energy for the carbon-side graphene is lower than for the Si-face graphene.
At this stage it is premature to speculate from which graphene layers (the
highly doped layer close to the substrate or nearly undoped top layers) this
peak stems and to conclude on its physical origin.

\subsection{Effective medium approximation for isolated disk-shaped quantum dots}\label{EMA}

The effective medium approximation is a standard approach to calculate
optical properties of inhomogeneous media. In the case of isolated quantum
dots of a two dimensional electron gas, the Maxwell-Garnett effective medium
model results in the following relation\cite{AllenPRB83S}:
\begin{equation}
\sigma_{\mbox{\small eff}}(\omega) = \frac{f\, \sigma(\omega)}{1 + C\, i \sigma(\omega)/(d \omega \kappa)}.
\label{ema}
\end{equation}
\noindent Here $\sigma(\omega)$ is the intrinsic conductivity of the electron
gas, $\kappa = (\epsilon_{1} + \epsilon_{2})/2$ is the average dielectric
function of the surrounding media, $f$ is the filling factor, $d$ is the dot
diameter and $C$ is a geometrical parameter related to the depolarization
factor of the dots. For the circular shape, the exact value for $C$ is
$3\pi^2/2$ \cite{LeavittPRB86S,MikhailovPRB96S}.

The optical conductivity of a homogeneous 2D electron gas or of highly doped
graphene in magnetic field, assuming constant scattering, is given by:
\begin{equation}
\sigma_{\pm}(\omega) = \frac{n e^2}{m}\frac{i}{\omega \mp \omega_c + i \gamma},
\label{sigmaplusmin}
\end{equation}
\noindent where $n$ is the density, $m$ is the mass, $\omega_{c}$ is the
cyclotron frequency and $\gamma$ is the scattering rate of the charge
carriers. Using eq. \ref{ema}, we obtain the following effective
conductivity:
\begin{equation}
\sigma_{\mbox{\small eff},\pm}(\omega) = \frac{f n e^2}{m}\frac{i}{\omega \mp \omega_c + i \gamma -
\omega_{0}^2/\omega},
\label{sigmaeff}
\end{equation}
\noindent which has a Lorentzian shape with a resonant energy given by:
\begin{equation}
\omega_{0}^2 = \frac{C n e^2}{m d \kappa}.
\label{Eqomega0}
\end{equation}
\noindent One can see that the plasmon spectral weight, $D = \int_0^\infty
\mbox{Re}\sigma_{xx}(\omega)d\omega = f \pi n e^2/2m$,  in eq. \ref{sigmaeff}
is reduced with respect to the Drude weight in eq. \ref{sigmaplusmin} by the
filling factor.

\subsection{Spectral weight and scattering}\label{SWGamma}

\begin{figure}
    \includegraphics[width=12cm]{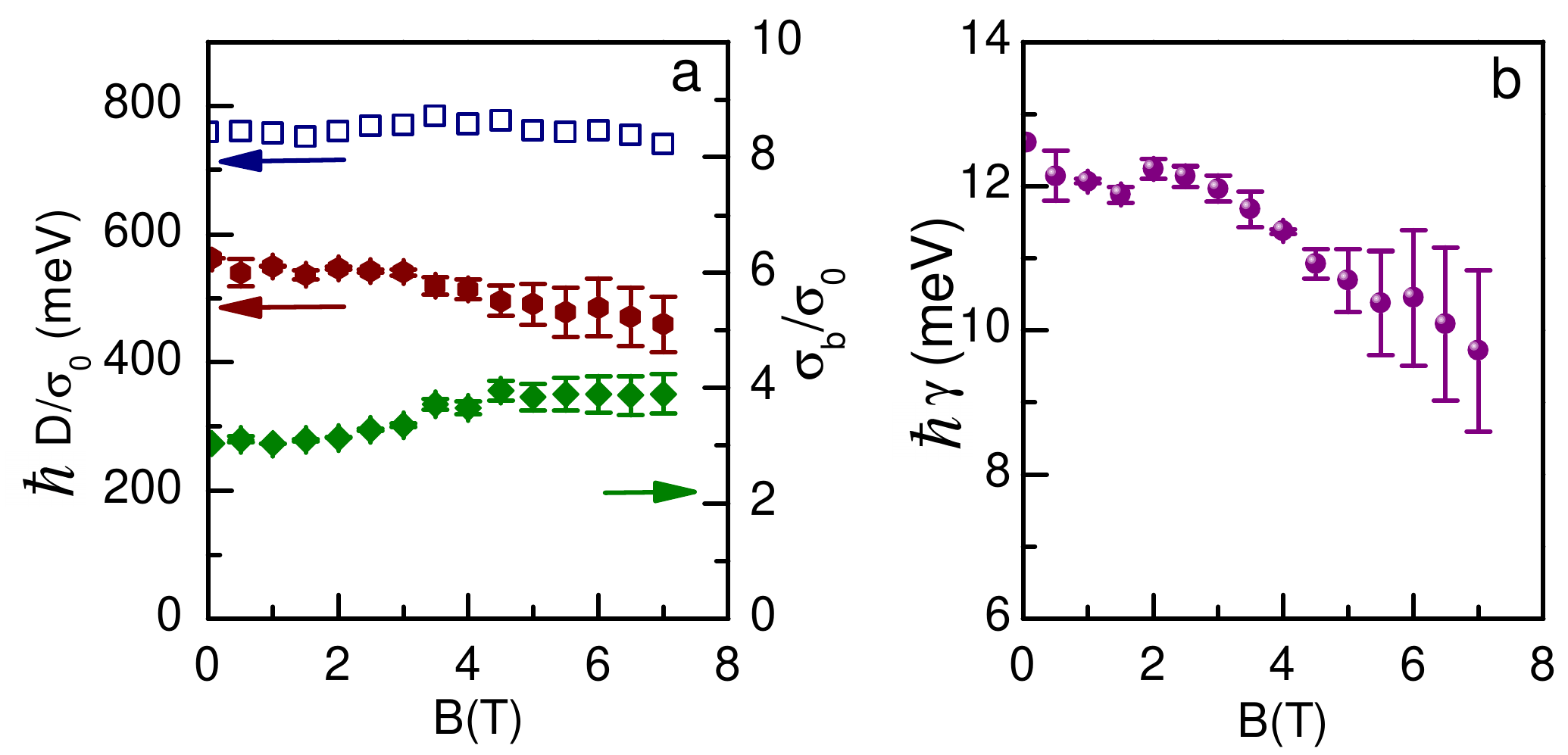}\\
    \caption{\textbf{Supporting Information.} (a) The magnetic field dependence of the plasmon
    spectral weight $D$ (hexagons), and the background term $\sigma_b$ (diamonds).
    (b) The field dependence of the broadening $\gamma$ (circles) of the plasmon resonance. $D$, $\sigma_b$ and $\gamma$ are obtained
    by fitting experimental data as described in the main text. Also shown in (a) is the
    integrated experimental optical conductivity $D_{int}$ (squares).}
    \label{Fig2}
\end{figure}

Figure \ref{Fig2}a shows the spectral weight of the plasmon resonance peak
and the background conductivity as a function of magnetic field, obtained by
fitting $\sigma_{xx}$ and $\sigma_{xy}$ to the effective medium model, as
described in the main text. Although the parameter $D$ shows a small decrease
with magnetic field, there is no reason to associate this with a true
decrease of the intraband spectral weight. In fact, the spectral weight
defined as the integrated experimental optical conductivity (from 2 to 85
meV), $D_{int} = \int\mbox{Re}\sigma_{xx}(\omega)d\omega$, shown in Figure
\ref{Fig2}a as open square symbols, is roughly constant. This is related to
the fact that the background conductivity shows a slight increase with field,
which effectively compensates the decrease of $D$. One can speculate about
the possible origin of the background term. It may be partially due to an
experimental uncertainty in the transmission normalization procedure. It is
not excluded, however, that $\sigma_b$ mimics a transfer of spectral weight
from the main peak to higher energies, due to electron-electron
\cite{PeresPRB06S}, electron-phonon interactions \cite{ParkNanoLett08S,
CarbottePRB10S} or excitations of higher harmonic plasmon
resonances\cite{LeavittPRB86S, MikhailovPRB96S, NikitinArxiv11S}. An
interesting possibility is that the negative charge removed from graphene
forms a weakly conductive layer in SiC, which causes a highly overdamped
Drude peak seen as a background in the studied spectral range. Notably, a
certain optical background below the onset of interband absorption was
present in exfoliated monolayer graphene on Si/SiO$_2$\cite{LiNaturePhys08S}.

We also observe a decrease of the plasmon scattering rate, as shown in Figure
\ref{Fig2}b. A plausible explanation is that the apparent scattering rate is
determined not only by the intrinsic electronic scattering, but also by a
statistical distribution of the resonance energy due to a varying size of
homogeneous regions and anisotropy. At high fields the position of the peak
is less influenced by the plasmon energy and the effect of statistical
distribution is weaker.

\end{document}